# Direct-written polymer field-effect transistors operating at 20 MHz

Andrea Perinot[1,2], Prakash Kshirsagar[3], Maria Ada Malvindi[3], Pier Paolo Pompa[3,4], Roberto Fiammengo[3,*], Mario Caironi[1,*]

[1] Center for Nano Science and Technology@PoliMi, Istituto Italiano di Tecnologia, via Giovanni Pascoli 70/3, Milan, Italy

[2] Dipartimento di Fisica, Politecnico di Milano, Piazza Leonardo da Vinci 32, Milano, Italy

[3] Center for Biomolecular Nanotechnologies@UniLe, Istituto Italiano di Tecnologia (IIT),Via Barsanti, 73010 Arnesano, Lecce, Italy

[4] Istituto Italiano di Tecnologia, Via Morego 30, 16163 Genova, Italy

[*] mario.caironi@iit.it, roberto.fiammengo@iit.it

**Abstract**

Printed polymer electronics has held for long the promise of revolutionizing technology by delivering distributed, flexible, lightweight and cost-effective applications for wearables, healthcare, diagnostic, automation and portable devices. While impressive progresses have been registered in terms of organic semiconductors mobility, field-effect transistors (FET), the basic building block of any circuit, are still showing limited speed of operation, thus limiting their real applicability. So far, attempts with organic FET to achieve the tens of MHz regime, a threshold for many applications comprising the driving of high resolution displays, have relied on the adoption of sophisticated lithographic techniques and/or complex architectures, undermining the whole concept.

In this work we demonstrate polymer FETs which can operate up to 20 MHz and are fabricated by means only of scalable printing techniques and direct-writing methods with a completely mask-less procedure. This is achieved by combining a fs-laser process for the sintering of high resolution metal electrodes, thus easily achieving micron-scale channels with reduced parasitism down to 0.19 pF mm$^{-1}$, and a large area coating technique of a high mobility polymer semiconductor, according to a simple and scalable process flow.

**Introduction**

Printed organic electronics has become a widespread research field in the recent years, thanks to the increased compatibility of the fabrication techniques with flexible, low-cost substrate materials, which results in the tangible prospect of economically convenient mass production of distributed electronics.[1-4] Notable demonstrations have been proposed for light detectors,[5,6] sensors,[7,8] transistors,[9] and on their integration into truly fully-printed opto-electronic circuits,[10-14] setting a step forward in the adoption of this technology into real life. On the other hand, the optimization of some desirable figures of merit still has to be appropriately approached to enable the implementation of a wide range of real applications. To this goal, it is of key importance to enhance the performance of the single organic transistors constituting the basic building block of almost any electronic circuitry.

Primarily, a transistor operation frequency suitable for more demanding applications must be guaranteed, *e.g.* several MHz at least in the case of addressing electronics for high-resolution flexible displays [15], wireless communication devices,[16-18] and RFID-based item-tracking systems.[19,20] Yet, in the majority of works, AC operation is either not assessed or considered a secondary goal compared to DC performance, and it is only marginally developed and analyzed.

It is well established that in typical organic field-effect architectures the optimization of high-frequency operation, besides requiring the adoption of semiconducting materials guaranteeing high effective mobility[21,22] and of dielectric layers with high relaxation frequency and yielding interfacial properties promoting efficient charge transport,[23,24] is largely determined by the reduction of the device channel length and of gate parasitic capacitance.[25] Nonetheless, this operation increases the influence of the charge injection limitations on the overall behavior of the device.[26-28] Fulfilling such requirements with a mask-less and fully solution-based process flow, albeit challenging, is highly desirable for the development of cost-effective technologies for wearable, portable and distributed applications. However, while promising results in AC operation of organic Field-Effect Transistors (FETs) have been reported, with a maximum operating frequencies as high as 27.7 MHz,[29] record values have been so far mainly achieved by adopting photolithographic steps[30-33] and/or single crystal semiconductors.[34] To date, one order-of-magnitude lower operating frequencies have been demonstrated when mask-less, scalable techniques were used, with a maximum of 3.3 MHz. [25,35-37]

In this work, we used a complete mask-less approach for the development of fully solution-processed polymer FETs capable of operating at frequencies as high as 20 MHz. In particular, we demonstrated the successful combination of printing techniques, such as large-area bar-coating,[38] and digital fs-laser sintering. [39-41] The latter is a direct writing technique for the fine patterning of a specifically formulated silver nanoparticle (Ag-NP) ink that allowed the fabrication of high-resolution conductive features down to a minimum lateral size of ~1.3 µm, to be used as electrodes for high performance polymer FETs.

**Results**

**Ag-NP ink preparation:** We have specifically developed a silver ink for the fs-laser sintering process, prepared starting from the synthesis of AgNPs with narrow size distribution, using a mixture of sodium citrate and tannic acid as reducing and stabilizing agents in aqueous conditions.[42] Other synthesis protocols have been recently proposed using tannic acid for controlling nanoparticle size either in combination with sodium citrate[43] or by careful control of the reaction mixture pH.[44] However, the reported conditions appear to be unsuitable for the preparation of AgNPs in such large amounts as required for conductive ink applications. We have, therefore, conducted an optimization of the reaction conditions and in particular temperature, reagents ratio and order of addition of the reagents, to allow the preparation of 23.8 ± 4.0 nm AgNPs (Figure 1, a and b) with a production scale-up factor between 150 and 300 times and yet of comparably narrow size distribution as those previously reported in the literature.[42-44] In a typical synthesis, our method yields 20 times more AgNPs also with respect to the approach recently reported by Zhang and coworkers.[45] After synthesis, the AgNPs were concentrated by centrifugation and coated by polyvinylpyrrolidone (PVP) of relatively low molecular weight (10 kDa). The choice of this polymer allowed us to increase the stability and processability of the AgNPs colloidal solution, while keeping thin the organic layer on the nanoparticle surface, a desirable property when aiming at conductive inks. The final AgNPs-based conductive inks have silver content of approx. 7% in water as the dispersion medium.

**Fs-laser sintering process characterization:** The AgNP ink was exploited to fabricate the device electrodes, following the process flow schematically illustrated in Fig. 1c. First, a uniform film of the conductive ink is deposited onto a substrate by spin-coating. Then, a femtosecond pulsed laser beam ($\lambda$ = 1030 nm, 67 MHz repetition rate) locally heats the nanoparticle film and patterns the desired geometry of conductive features which are ~70 nm thick. As a final step, the unprocessed areas of the film are washed away using water, and only the conductive high-resolution patterns are left on the substrate. Differently from what previously reported in literature,[41] our fs-sintering process does not require any ink encapsulation prior to laser processing, thus largely simplifying the process and making it potentially scalable.

To properly control the metallic electrodes fabrication process and to determine the optimal conditions for their integration into FETs, we performed a thorough study and related the fabrication parameters, such as power (in the range 5 – 30 mW, corresponding to the range 0.8-4.9 mW/µm$^2$), scanning speed (in the range 0.05 – 1 mm s$^{-1}$) and magnification power of the optics (20X, 50X or 100X), to the dimensions and conductivity of the sintered patterns. We fabricated 190 µm-long single lines varying said process parameters and measured their lateral width and electrical resistance. In Fig. 1e we relate the obtained line width with the beam power for different scanning speeds: an approximately linear dependence can be identified in the range from 5 to 30 mW, yielding linewidths from a minimum of 1.3 to 5.3 µm. A similar trend is highlighted also when testing the line width versus beam power varying the optics (0.1 mm/s, Fig. 1d).

The conductivity of the laser-sintered lines was compared to the bulk-Ag conductivity for different beam power and scanning speed (Fig. 1f). A general increase of the conductivity is observed when increasing the beam power or slowing down the scanning speed, and its value ranges between 4.5 and 21 kS/cm, which corresponds to a range from 0.7 to 3.3 % of bulk silver conductivity. This range is in line with the typical conductivities obtained with more established direct-writing techniques, such as inkjet printing.[46] This validates the suitability of this process and ink for the fabrication of high-resolution conductive patterns for organic electronics applications.

We then employed fs-laser sintering for the fabrication of the source and drain bottom contacts of polymer FETs, in order to benefit both from the reduction of the channel length and of the width of the contacts. Previous examples where fs-laser sintered contacts were used for organic FETs are reported in.[41,47] Such examples either did not integrate high-resolution, reduced parasitic capacitance electrodes,[41] or adopted more expensive Au inks.[47] Both reported only bottom-gate, bottom-contacts architectures, which, while more easily accessible, are not as performing as staggered, top-gate devices for injection in downscaled channels. Indeed, very limited performance in terms of mobility were achieved ($\mu$ lower than 10$^{-2}$ cm$^2$/Vs for channel lengths longer than 5 µm). Therefore, previous reports do now answer the question whether fs-laser sintering is a candidate for the fabrication of high-frequency, direct-written polymer transistors, which is the main goal of the present work.

**OFETs fabrication and DC characterization:** We fabricated high-resolution short- and long-channel *n*-type OFETs in a staggered, top-gate configuration (Fig 2d), which is known to yield better charge injection properties from the contacts.[26] The fabrication process flow is illustrated in Fig. 2a. For the contacts, following the optimization process described above, the chosen laser processing parameters were set to 12 mW beam power (1.9 mW/µm$^2$), 50X optics, 0.5 mm/s scanning speed. A typical AFM image of a fabricated contact is shown in Fig. 1g. We adopted the semiconducting co-polymer poly[N,N'-bis(2-octyldodecyl)-naphthalene-1,4,5,8-bis(dicarboximide)-2,6-diyl]-alt-5,5'-(2,2'-bithiophene), P(NDI2OD-T2) (Fig. 2c), which was deposited through bar-coating, a scalable large-area printing technique.[2,38] A 500 nm-thick PMMA layer was deposited through spin-coating, and finally we selectively inkjet-printed PEDOT:PSS over the channel and contact area to create the top gate contact. We fabricated devices with different channel lengths $L$ of 1.75, 5.1 and 21.6 µm. The contact width $L_{ov}$ was 3.0 ± 0.5 µm and the channel width $W$ was 800 µm, both constant among all devices. A micrograph of the realized pattern, prior to semiconductor deposition, is shown in Fig. 2b. In Fig. 2e and 2f, we show the measured transfer curves for two long channel devices ($L$ = 5.1 µm, red line, and $L$ = 21.6 µm, blue line) and for a short-channel device ($L$ = 1.75 µm). We summarize in Table 1 the extracted apparent mobilities for the full set of fabricated OFETs.

The devices exhibit proper *n*-type operation and in the linear regime ($V_d$ = 5 V) the drain current scales accordingly with the reduction of the channel length. The correct scaling is confirmed by the extracted values of linear charge mobility, which remains comparable (in the range 0.1-0.2 cm$^2$V$^{-1}$s$^{-1}$) while the channel length is scaled by one order of magnitude (Figure S2), suggesting good charge injection behavior at the semiconductor/contact interface. We performed a first approach estimation based on the TLM method, which can slightly underestimate the contact resistance for the semiconductor in use,[48] and found a normalized contact resistance $R_cW = 7.3$ kΩcm at $V_g = 40$ V (figure S1), indicating that the scaled electrodes fabricated with our method inject and collect charges effectively.

In the saturation regime we extracted good effective mobilities of 0.37 cm$^2$V$^{-1}$s$^{-1}$ and 0.74 cm$^2$V$^{-1}$s$^{-1}$ for the devices with $L = 21.6$ µm and $L = 5.1$ µm respectively ($V_g = V_d = 40$ V), and a high effective mobility of 0.82 cm$^2$V$^{-1}$s$^{-1}$ ($V_g = 22.5$ V; $V_d = 40$ V) in the case of the shortest channel length of 1.75 µm. In the latter case, we observed a breakdown of the device at currents above 0.9 µA/µm when $V_d = 40$ V and $V_g > 25$-30 V, so we limited the driving voltage accordingly, below the onset of breakdown. We remark how the measured effective mobilities in the saturation regime increase when shortening the channel length, in contrast with the behavior exhibited by the linear regime. In fact, we identified a gate dependence of the charge mobility in the saturation regime whose effect is amplified at shorter channel lengths (Figure S3). This behavior is not unique to our work, it is general for the semiconductor in use, it characterizes several high mobility donor-acceptor copolymers recently reported and it can be explained by an effect of the lateral field on injection and/or transport.[49-51]

**OFETs AC characterization and high-frequency operation:** The successful adoption of high resolution fs-laser sintered contacts in polymer FETs allowed us to assess the impact on the maximum operational frequency of the fabricated devices. First, it is worth recalling that, at a first approximation, the final limitation to the speed of a FET is determined by the charge carrier transit time across the channel: $t_{tr} = \frac{L}{v} = \frac{L}{\mu E} = \frac{L^2}{\mu V}$, where $v$ is the carrier velocity, $E$ is the lateral electric field across the channel and $V$ is the drain-to-source voltage. However, additional parasitism intervenes to reduce the actual maximum operational frequency well below this limit. In particular, the overlap capacitances insisting between gate and source/drain electrodes are the most critical in the low-resolution architectures (*e.g.* tens of *µm* for inkjet-printed contacts) typically adopted in organic FETs fabrications. A more appropriate figure of merit to describe the maximum operational frequency of a FET is the transition frequency $f_t$, corresponding to the frequency at which the total AC gate, or "input", current $i_g$ becomes equal to the AC drain, or "output" current $i_d$. $f_t$ depends on the device transconductance $g_m$ and on the gate/source and gate/drain capacitances $C_{gs}$ and $C_{gd}$ according to the expression[25]

$$f_t = \frac{g_m}{2\pi\,(C_{gs}+C_{gd})} \quad (1)$$

In order to extract $f_t$, we measured separately $C_{gs}$, $C_{gd}$ and $g_m$ as a function of frequency for our devices. Details on the extraction methodology can be found in the SI. Source and drain capacitances were measured both by operating the FETs at $V_g$ = -15 V and $V_d$ = -15 V (off-state regime), when we expect to probe only the physical overlap capacitances, and at $V_g$ = 25 V and $V_d$ = 25 V (on-state, saturation regime), when we expect an increased capacitance, due to the accumulated channel region. In Fig. 3b we show the measured $C_{gs}$ in both regimes for a device with $L$ = 1.75 µm: in off-state regime a capacitance of 0.10 pF is measured, which is in very good agreement with the theoretical value of 0.096 pF, calculated for a parallel-plates capacitor model with an area corresponding to the geometrical overlap between the source and gate. When accumulating the channel, we measured a capacitance increase of 0.03 pF, which is attributed to the addition of the channel area to the electrodes overlap capacitance. We measured the gate-source capacitance in the saturation regime for different channel lengths and linearly fitted the measured data (Figure S4). The intercept of the fitted line at $L$ = 0 µm, identifying the capacitance attributed to the overlap between the contacts, returns 0.085 pF, consistently with the measured and theoretically calculated values. The slope of the fitted line, representing the accumulated channel capacitance per unit channel length, returns 0.026 pF/µm. If $C_{diel}$ is the dielectric capacitance per unit area, this slope can be theoretically estimated as $\frac{C_{ch,sat}}{L} = \frac{2}{3} W C_{diel} = 0.034$ pF/µm,[52] which is in good agreement with the fitting. We also measured the gate-drain capacitance in the saturation regime: $C_{gd}$ is equal both in accumulation and in the off-state, corresponding to 0.21 pF for the shortest channel length and consistently with the fact that, in this regime, we are only probing the gate/drain overlap capacitance. The good agreement between the measurements and theoretical value of $C_{gd}$ is obtained also for longer channel lengths. All capacitance data, both for $C_{gd}$ and $C_{gs}$, have been summarized in Figure 3c. We measured the transconductance of the short-channel device in the range 1 kHz - 2 MHz. Differently from the ideal behavior, a slight roll-off of this figure with frequency can be identified which reduces $g_m$ from 52.8 µAV$^{-1}$ at 1 kHz to 42.8 µAV$^{-1}$ at 2 MHz (Fig. 3d, black solid line). The maximum frequency limitation of the setup prevented the measurement of the transconductance over 2 MHz, and we determined the transition frequency by linearly extrapolating the measured trends of $g_m$ and $C_g$. In Fig. 3d we show the measurement and the determined $f_t$ = 20 MHz. A simple calculation from the measured $g_m$ at 1 kHz and the measured $C_g$ according to (1), would return an $f_t$ = 24.7 MHz. Recalling the discussion above, we validate our result by comparing the frequency obtained by AC measurements with the ultimate limit defined by the transit time of the carriers across the channel: for our shortest-channel device we obtain $t_{tr}$ = 0.9 ns corresponding to an ideal cutoff frequency of $f_{tr} = \frac{1}{2\pi t_{tr}}$ = 170 MHz. The extrapolated $f_t$ is well below the calculated cut-off frequency due to the carrier transit time, thus our estimation is not undermined by the extrapolation method.

**Discussion**

In Table 2 we compare the result obtained in this work with previous works in which the transition frequency has been explicitly measured, and we report, along with $f_t$, other relevant figures of merit (i.e. operational voltage and effective mobility) and fabrication details of the devices. Please note that the reported mobility values are effective parameters, largely influenced by the specific architecture adopted, as well as by the semiconductor processing method and microstructure. To the best of our knowledge, our demonstration reaches the highest $f_t$ reported for an *n*-type printed polymer device so far,[36] and of the same order of magnitude of the highest reported in the organic electronics literature in absolute terms, including devices realized through process flows including photolithography or evaporation steps. Overall, we demonstrate the fastest organic FET fabricated to date without the use of any mask and by combining only printing and direct-writing processes.

The strong enhancement of this figure of merit with respect to previous works dedicated to maskless fabrication of organic electronics, limited to 3.3 MHz in the best case,[36] is ascribed to multiple contributions. The drastic improvement in the resolution for the definition of the OFETs source and drain contacts, owing to the use of fs-laser sintered high-conductive AgNP electrodes, is combined with the bar-coating of an optimized semiconductor morphology[2] enabling enhanced effective field-effect mobility, also thanks to good charge injection properties of such electrodes. The aforesaid fabrication strategy improves the patterning resolution of a typical direct-writing technique, e.g. inkjet printing, by more than 10 times [4,53] and obeys the multiple role of reducing both the transistor channel length and the parasitic capacitance present between these electrodes and the gate. Moreover, the proposed process retains additional room for improvement in terms of reduction of the parasitic capacitance, for example with the integration of a self-alignment technique for the patterning of the gate electrode.[54]

In addition to properly controlling the transistor features that are critical for high-frequency operation, the selected fabrication techniques have strong potential for scalability to mass production, featuring both screen-to-screen and roll-to-roll compatibility.[55] The fs-laser process necessitates only of a limited beam power (as low as 10 mW), hence throughput can be enhanced through parallelization of different beams, independently driven (possibly by galvo-scanners), generated from the same laser source. A further possibility for the improvement of the throughput is the adoption of Spatial Light Modulators, which may enable the possibility to write sections of a device or a circuit in a single step. Additionally, we recall that, in a practical application, only a fraction of the features needs to be defined with a high resolution, so that the effective throughput can be further enhanced. In perspective, these features enable the applicability of the proposed process to large-volumes manufacturing at reduced cost of circuits incorporating polymer transistors operating beyond 20 MHz, therefore widely expanding possible applications of cheap flexible electronics.

**Methods**

**General (NP-ink synthesis):** All reagents were purchased from Sigma-Aldrich, and used without further purification: silver nitrate (nr. 209139, ACS reagent, ≥99.0%), tannic acid (nr. 403040, ACS reagent), sodium citrate tribasic dehydrate (nr. 71402, BioUltra, ≥99.5%), polyvinylpyrrolidone (nr. PVP10, average m.w. 10 kDa). All glassware employed for AgNP preparation was cleaned with conc. $HNO_3$ and rinsed with plenty of water. Ultrapure deionized water (Millipore purification system, 18.2 MΩ cm) was used for the preparation of all aqueous solutions. All solutions used for nanoparticle preparation were filtered through a 0.2 µm membrane filter (Whatman, cellulose acetate). AgNPs were characterized by Dynamic Light Scattering (DLS) (Zetasizer Nano ZS, Malvern Instruments). AgNP diameter was measured via TEM (JEOL JEM-1011 transmission electron microscope operating at an accelerating voltage of 100 kV). UV/Vis spectra and fluorescence measurements were carried out using a TECAN Infinite M200 Pro plate reader. The concentration of silver in the inks was determined via ICP-OES (Agilent 720 ICP-OES).

**Nanoparticle synthesis:** Silver nitrate (1.274 g, 7.50 mmol) was dissolved in 1.45 L of water. The solution was heated up with efficient stirring and protected from light until it started to reflux followed by quick addition of 50 mL of a freshly prepared solution of sodium citrate (0.302 M, 15.1 mmol) and tannic acid ($7.35 \times 10^{-4}$ M, $3.68 \times 10^{-2}$ mmol) in water. Accordingly, the concentrations of reactants in the reaction mixture were: [Ag]= $5.00 \times 10^{-3}$ M, [Na citrate]= $1.00 \times 10^{-2}$ M, and [tannic acid]=$2.45 \times 10^{-5}$ M. The solution rapidly turned dark indicating nanoparticle fomation, but heating was continued for additional 30 minutes. Heating was then removed and the reaction mixture was allowed to reach room temperature overnight. The AgNPs were concentrated by centrifugation (20 min at 10000×g).

**PVP-coating of AgNPs:** The concentrated AgNPs were redispersed in water containing Polyvinylpyrrolidone (PVP, average m.w. 10 kDa) to afford a 50 mL mixture with [PVP]=$6.4\times10^{-3}$ M. This mixture was stirred for 30 min. and then centrifuged 1h at 10000×g. Most of the supernatant was removed (> 40 mL) and the coating procedure was repeated a second time with overnight incubation. The PVP-coated AgNPs were concentrated to approx. 10 mL by centrifugation (1.5 h at 10800×$g$) and then washed with water (2×50 mL) by centrifugation and resuspension. The aqueous AgNP solution was filtered over a 0.2 µm cellulose acetate filter before the last centrifugation. The volume of the final PVP-coated AgNP concentrate was approximately 4 mL. Each resuspension step after centrifugation included 20 min. incubation in an ultrasonic bath.

**Ink preparation:** 2 mL of aqueous PVP-coated AgNP concentrate were added of 16 mL of water. The AgNP suspension was centrifuged (1 h, 20 °C at 10800×$g$), the supernatant removed and the AgNP pellets taken up in 3 mL of water. For this ink: 4.3 mL, 67.1 g/L Ag, DLS: 26.2±0.2 nm, PDI 0.532, $\lambda_{max}$=399 nm (in $H_2O$), TEM: 23.8 ± 4.0 nm

**Femtosecond laser sintering setup:** The laser setup consists of a commercial laser source ( LightConversion PHAROS, based on Yb:KGW as active medium) which generates ~80 fs-long laser pulses with a repetition rate of 67 MHz, $\lambda$ = 1030 nm and maximum output power of 2 W. Before reaching the sample, the beam is conditioned through an optical path which includes a software-controlled attenuator and a focalizing objective (Mitutoyo) lens whose magnifying power can be selected between 20X, 50X or 100X. The sample is positioned on a software-controlled moving stage (Aerotech ABL1000) capable of a maximum resolution of 0.5 nm and a maximum speed of 300 mm/s.

**Contacts fabrication:** Standard glass slides are used as a substrate. They are cleaned in an ultrasonic bath with deionized water, acetone, 2-propanol sequentially for 5 minutes each, then oxygen-plasma treated for 1 min (100 W), and finally heated on a hotplate at 60 °C for 5 minutes. The AgNP ink is then spincoated at 1000 rpm for 40 s. After laser processing, the sample is rinsed and sonicated for 1 min with deionized water.

**Conductive pattern characterization:** Optical images are taken with a Zeiss Axio Scope.A1 microscope (100X objective), AFM images are taken with an Agilent 5500 Atomic Force Microscope and SEM images with a JEOL JSM 6010LV. Electrical characterization is performed using an Agilent B1500A Semiconductor Parameter Analyzer.

**Organic FETs fabrication:** P(NDI2OD-T2) (purchased from Polyera) is dissolved in mesitylene at a concentration of 5 mg/ml. After fabrication of the bottom contacts via laser sintering, P(NDI2OD-T2) is deposited through bar-coating in air atmosphere, using the same process as described in [2]. Then poly(methyl methacrylate) (PMMA) is spun from n-butylacetate (concentration 80 mg/ml) at 1500 rpm for 1 minute. After dielectric deposition, the devices are annealed on a hotplate for 30 min at 80 °C for residual solvent removal. PEDOT:PSS (Clevios P Jet 700) is patterned over the contacts and channel area via inkjet (using a Fujifilm Dimatix DMP-2831). The devices are then annealed at 120 °C in nitrogen atmosphere for 12h.

**Electrical characterization:** The devices are measured in nitrogen atmosphere. Static characterization is performed via an Agilent B1500A Semiconductor Parameter Analyzer. Frequency performance was measured using a custom setup which includes an Agilent ENA Vector Network Analyzer and an Agilent B2912A Source Meter. More details on the setup can be found in Supplementary Information.

**Acknowledgements**

The authors are thankful to: Krishna Chaitanya Vishunubhatla and Luigino Criante for the support with the femtosecond laser machining setup; Sadir Bucella for very useful help in the early stage of the research; to Michele Giorgio for assistance in the fabrication of part of the reported devices. This work has been financially supported by the European Research Council (ERC) under the European Union's Horizon 2020 research and innovation programme 'HEROIC', grant agreement 638059.

**Author Contributions**

A.P. and M.C. designed the experiments. A.P. performed the experiments on the Ag-NP ink deposition, fs-laser sintering process characterization, OFET fabrication and DC and AC characterization. P.K., M.A.M., P.P.P., R.F. synthesized the Ag-NP ink. All authors analyzed the data and contributed to the manuscript.

**Additional Information**

Supplementary information is available online or upon request. Correspondence should be addressed to R.F and M.C.

**Competing Financial Interests**

No competing financial interests of the authors are associated to this work

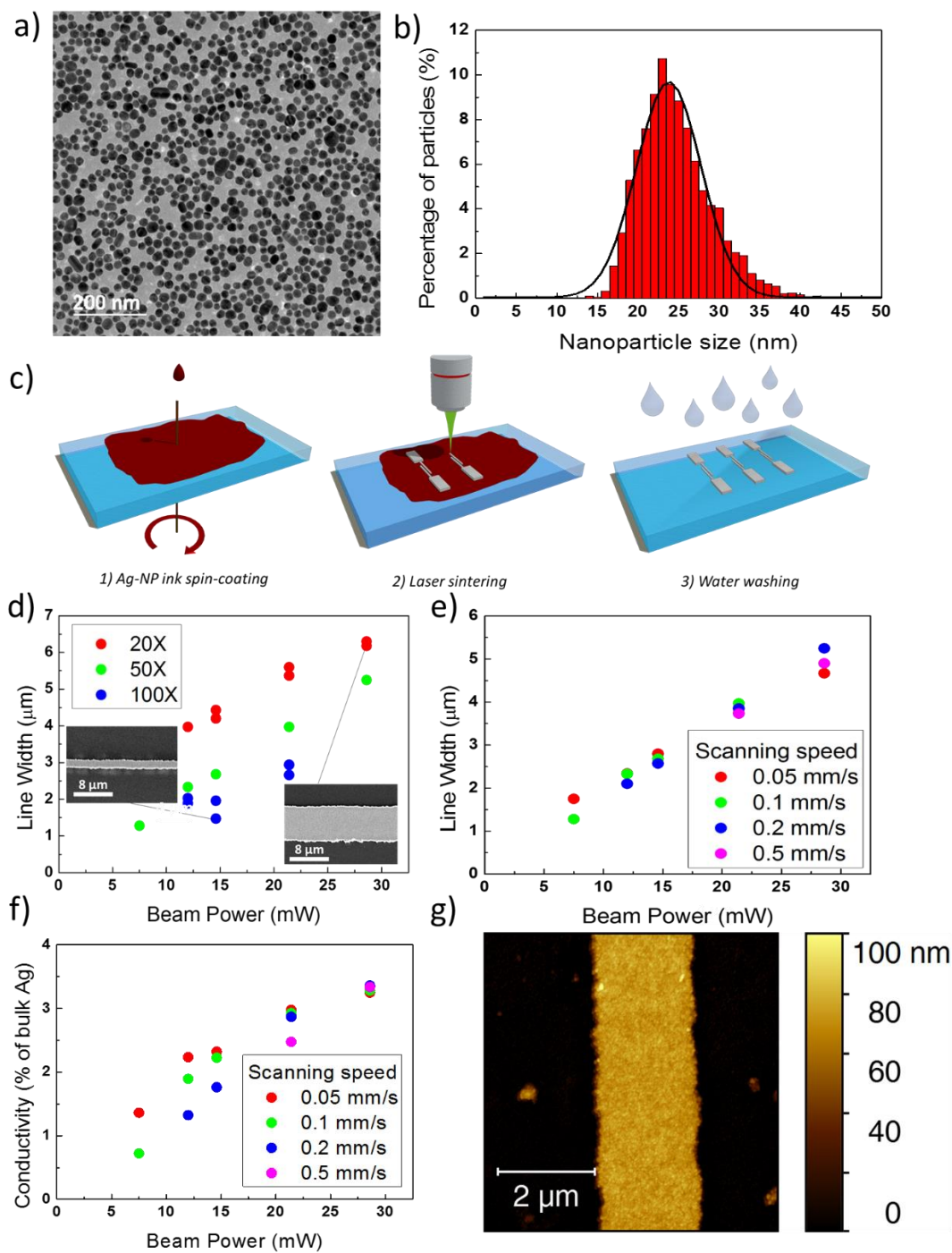

**Figure 1: Ag-NP ink properties and fs-laser sintering process.** a) TEM image and b) nanoparticle size distribution (~4500 particles)s of the AgNP ink, c) Scheme of the femtosecond laser sintering process flow, d) sintered line width vs. beam power for different optics and scan speed 0.1 mm/s (insets show SEM images of some of the lines), e) sintered line width vs. beam power for different scan speeds (50X optics), f) sintered line conductivity (relatively to bulk Ag conductivity) vs. beam power for different scan speeds (50X optics) and g) AFM image of a laser sintered Ag line.

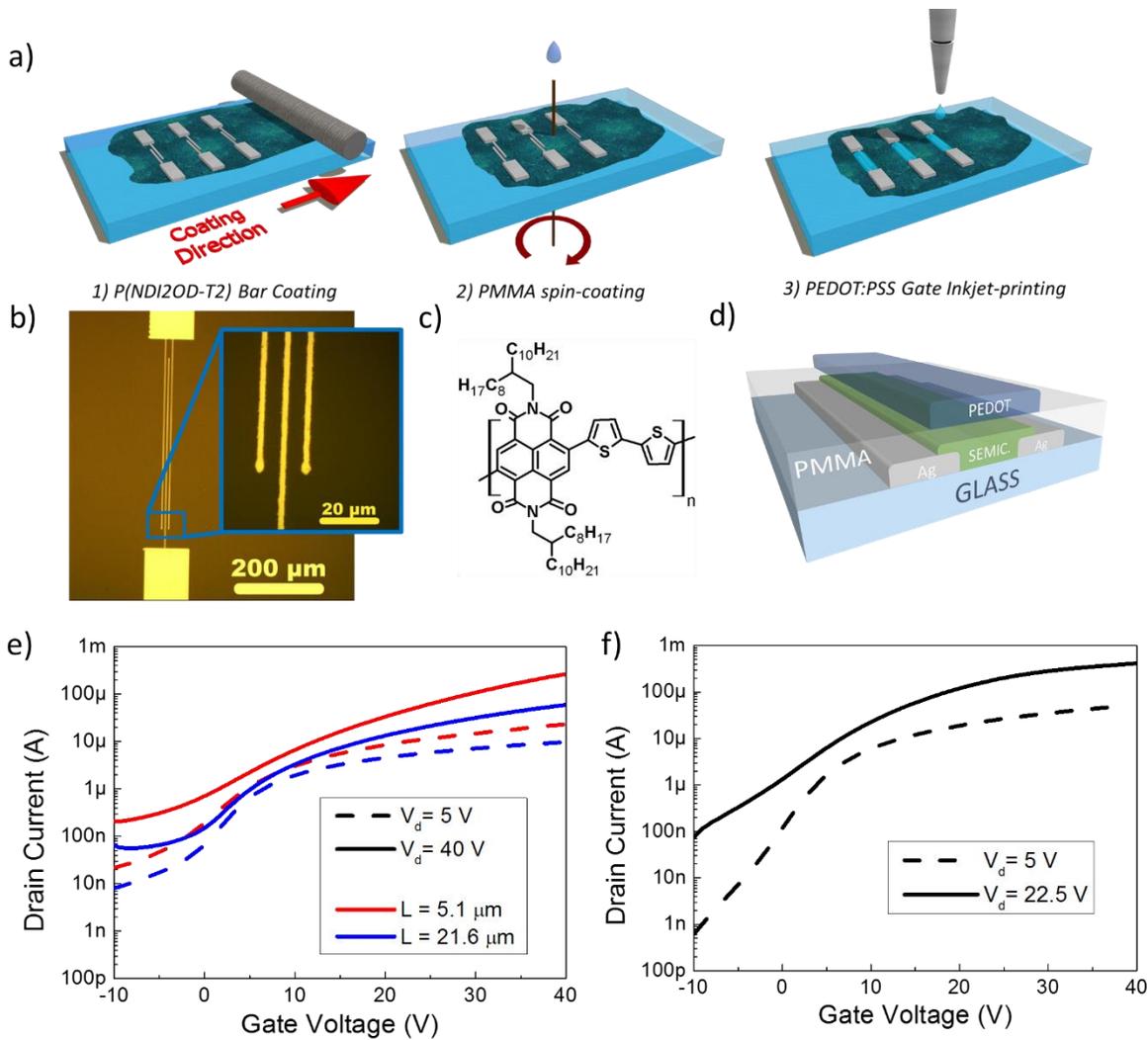

**Figure 2: OFET fabrication and DC characterization.** a) Sketch of the fabrication process, b) optical micrograph of the bottom source and drain electrodes realized through laser sintering, c) chemical structure of P(NDI2OD-T2), d) sketch of the realized OFET, and e) transfer characteristics of two devices with channel lengths *L = 5.1 µm* and *L = 21.6 µm* (*W = 800 µm*) and f) transfer characteristic of a short-channel device (*L = 1.75 µm, W = 800 µm*).

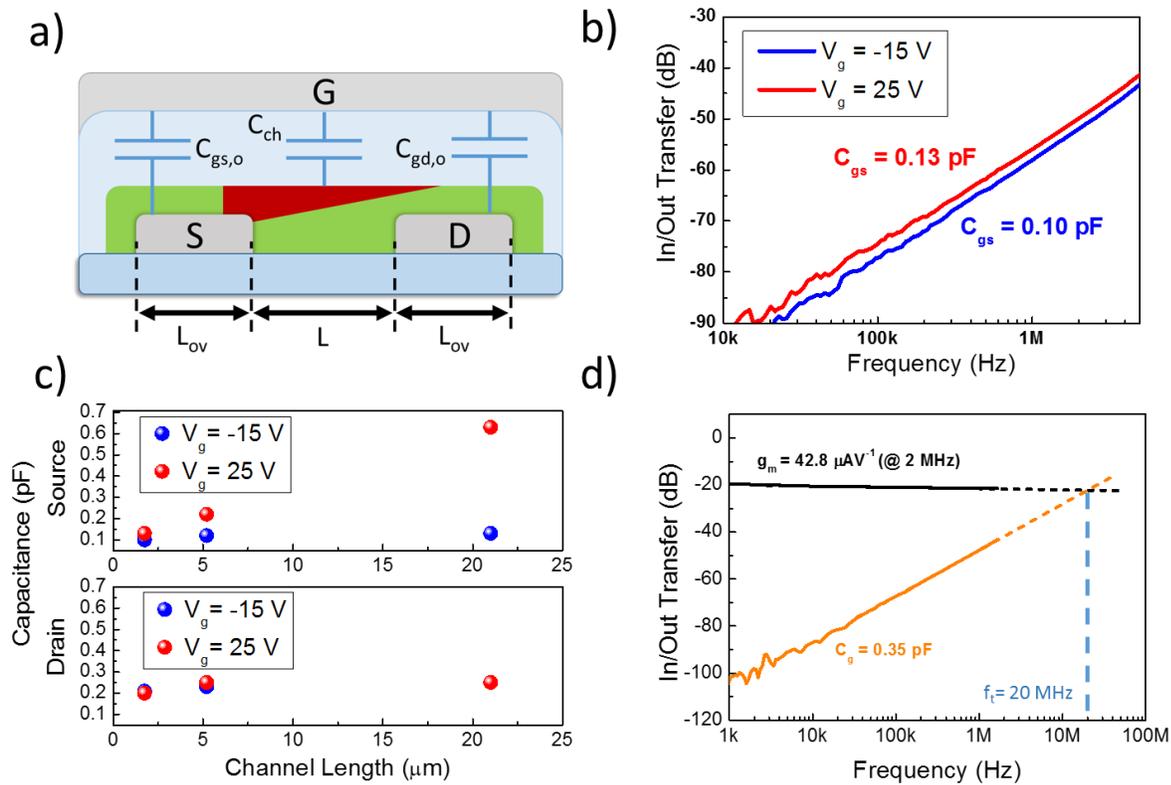

**Figure 3: AC characterization of high-frequency OFETs.** a) Sketch of the lateral view of the realized OFETs, highlighting the different capacitances insisting across the device electrodes, b) measured $C_{gs}$ in depleted-channel ($V_{gs} = -15\ V$, $V_{ds} = -15\ V$) and in accumulated-channel ($V_{gs} = 25\ V$, $V_{ds} = 25\ V$) regimes for an n-type device ($L = 1.75\ \mu m$), c) measured source and drain capacitances for different channel lengths in depleted- and accumulated-channel regimes and d) determination of the transition frequency for a device ($V_{gs} = 30\ V$, $L = 1.75\ \mu m$).

**Table 1.** Extracted mobilities for the laser sintered OFETs

| Channel length [μm] | Linear mobility [$cm^2V^{-1}s^{-1}$] | Saturation mobility [$cm^2V^{-1}s^{-1}$] |
|---|---|---|
| 1.75 | 0.13 | 0.82* |
| 5.1 | 0.14 | 0.74 |
| 21.6 | 0.22 | 0.37 |

*$V_g = 22.5$ V

**Table 2.** Properties and fabrication techniques for the OFETs with the highest reported transition frequencies (listed mobility values are to be considered effective parameters, largely depending on the specific device architecture).

| $f_t$ [MHz] | Ref. | Semiconductor | | Voltage [V] | Mobility [$cm^2V^{-1}s^{-1}$] |
|---|---|---|---|---|---|
| **With the use of photolithography/masks** | | | | | |
| 27.7 | 29 | C60 | Evaporated | 20 | 2.22 |
| 25 | 32 | Rubrene | Single Crystal | -15 | 10.3 |
| 20 | 30 | $C_{10}$-DNTT | Evaporated | -20 | 0.4 |
| 20 | 34 | DNTT | Evaporated | -15 | 0.44 |
| 19 | 33 | $C_{10}$-DNTT | Evaporated | -10 | 2.5 |
| 11.4 | 29 | Pentacene | Evaporated | -20 | 0.73 |
| 10 | 30 | $C_{10}$-DNTT | Sol. Crystallized[56] | -20 | 0.4 |
| 3.7 | 57 | DNTT | Evaporated | -3 | 0.7 |
| 1.5-2 | 23 | PBTOR | Spin-coated | -4 | 2.5 |
| **Direct-written electrodes** | | | | | |
| 20 | This work | P(NDI2OD-T2) | Printed (bar-coating) | 30 | 0.9 |
| 3.3 | 36 | P(NDI2OD-T2) | Printed (inkjet) | 30 | 0.005 |
| 2.8 | 36 | DPPT-TT | Printed (inkjet) | -25 | 0.024 |
| 1.6 | 25 | pBTTT | Spin-coated | -8 | 0.1 |